\documentclass[pra,amsmath,amssymb,floatfix,twocolumn,showpacs]{revtex4-1}
\usepackage{graphicx}
\usepackage{subfigure}
\usepackage[colorlinks=true,a4paper=true, pdfstartview=FitV,linkcolor=blue, citecolor=blue, urlcolor=blue]{hyperref}
\usepackage[all]{hypcap}
\usepackage{multirow}
\usepackage{times}
\usepackage{bm}
\begin{document}
\title{Dipolar Dynamics for Interacting Ultracold Fermions in a Trapped Optical Lattice}
\author{Jia-Wei Huo$^{1}$, Weiqiang Chen$^{1,2}$, U. Schollw\"ock$^{3}$, M. Troyer$^{4}$, Fu-Chun Zhang$^1$}
\affiliation{$^1$Department of Physics, The University of Hong Kong, Pokfulam Road, Hong Kong, China\\
$^2$Department of Physics, South University of Science and Technology of China, Shenzhen, Guangdong 518055, China\\
$^3$Department of Physics, Arnold Sommerfeld Center for Theoretical Physics, Ludwig-Maximilians-Universit\"at M\"unchen, D-80333 M\"unchen, Germany\\
$^4$Institute for Theoretical Physics, ETH Zurich, CH-8093 Zurich, Switzerland}
\date{\today}
\begin{abstract}
Using the time-dependent density matrix renormalization group (tDMRG) method, we calculate transport properties of an interacting Fermi gas in an optical lattice with a confining trap after a sudden displacement of the trap center. In the regime of attractive interactions, the dipolar motion after the displacement can be classified into underdamped oscillations and overdamped relaxations, depending on the interaction strength. These numerical calculations are consistent with experimental results. In the regime of repulsive interactions, we predict a revival of the oscillations of the center of mass when the interaction strength is increased. This unique feature can be considered as a dynamical signature for the emergence of a Mott plateau for an interacting trapped Fermi gas in an optical lattice.
\end{abstract}
\maketitle
\section{Introduction}
Ultracold atoms have proven to be an ideal platform to study a number of unsolved problems in quantum many-body physics. The unprecedented controllability of the system is experimentally achieved through optical lattices~\cite{Orzel2001} and Feshbach resonances~\cite{Inouye1998}. Optical lattices provide the possibility to manipulate the dimensionality of the lattice and the ratio of interaction to kinetic energy, while Feshbach resonances can be used to tune the interaction strength, crucial to many-body effects. All these developments provide a route to simulating many paradigmatic quantum models in strongly correlated systems and therefore are a novel approach for answering fundamental problems in many-body physics~\cite{Bloch2008}.

Driven by the ambitious goal of carrying out quantum simulations, experimentalists have begun to explore the underlying physics of strongly correlated fermions in the atomic approach. The Hubbard model is a simple but very good starting point for modeling the essence of interacting fermionic systems. In the meantime, great success has been achieved towards experimental realization of this model. The Fermi surface of degenerate Fermi gases has been observed in a three-dimensional optical lattice~\cite{Kohl2005}. Molecules from fermionic atoms have been created with a Feshbach resonance~\cite{Stoferle2006}, and a Mott insulator has been detected~\cite{Jordens2008,Schneider2008}. Although all these developments have paved the way for understanding fermionic superfluidity, transport properties are hard to access.

In one approach to transport phenomena, Strohmaier {\it et al.}~\cite{Strohmaier2007} carried out an experiment on an interacting Fermi gas in an optical lattice. They used the center of mass (COM) motion of an interacting Fermi gas to characterize its dynamical features. More specifically, the COM motion of the gases was monitored after a sudden shift of the external harmonic trap by a few lattice sites, giving rise to dipole excitations. Unlike similar experimental studies on bosonic atoms with repulsive interactions~\cite{Fertig2005}, they focused on the regime of attractive interactions and found that with increasingly attractive interactions weakly damped oscillation turns into a slow relaxational drift. 

A quantitative understanding of all these observations is a theoretical challenge due to the non-equilibrium nature of the process. In bosonic experiments~\cite{Fertig2005}, the dipolar motion due to the sudden shift reflects the dynamical excitations in the many-body system, and has stimulated considerable theoretical interest, involving semiclassical solutions~\cite{Pezze2004} and numerical diagonalization of the Bose-Hubbard Hamiltonian in a small system~\cite{Rey2004}. By using the time-dependent density-matrix renormalization group technique (tDMRG)~\cite{Schollwock2011,Verstraete2004,Vidal2003,Vidal2004,White2004,Daley2004}, simulations of the bosonic model have been successful in quantitative comparisons with the experimental data~\cite{Danshita2009,Montangero2009}. This indicates that tDMRG is an ideal tool in the strongly correlated regime in one spatial dimension even far away from equilibrium.

In this work, we present a comprehensive numerical simulations of the fermionic experiment by Strohmaier {\it et al.}~\cite{Strohmaier2007} in the attractive interaction regime, and extend the study to the repulsive interaction regime, which is also experimentally achievable. To take into account the full time dependence, we focus on one spatial dimension by using exact diagonalization and tDMRG. Our approach is similar to that of Okumura {\it et al.}~\cite{Okumura2010S949} who present results for the same system, but we go far beyond this short note by analyzing the results and explaining the qualitative and quantitative features and studying the repulsive case. In our tDMRG simulations, up to 400 states are kept in the reduced Hilbert space, and a second order Trotter decomposition is used in the time evolution. 

\section{Model}
Ultracold fermions in a one-dimensional optical lattice can be described by the Hubbard model~\cite{Jaksch1998,Esslinger2010}
\begin{multline}
  \hat{H}(t)=-J\sum_{i,\sigma}(\hat{c}_{i+1,\sigma}^\dag\hat{c}^{}_{i,\sigma}+ \text{H.c.})+U\sum_i\hat{n}_{i,\uparrow}\hat{n}_{i,\downarrow}\\
  +\Omega\sum_{i,\sigma}\hat{n}_{i,\sigma}[i-i_0+\Theta(t)\Delta x]^2.
\end{multline}
The first term describes the tunneling of fermions between nearest neighboring sites, where $J$ denotes the tunneling matrix element between adjacent lattice sites and $\hat{c}_{i,\sigma}$ the fermionic annihilation operator on site $i$ in the ``spin'' state $\sigma$ ($\uparrow$ or $\downarrow$). The second term is the on-site interaction with strength $U$, where $\hat{n}_{i,\sigma}\!\!=\!\!\hat{c}_{i,\sigma}^\dag\hat{c}^{}_{i,\sigma}$. The last term models the additional confinement of the harmonic trap with curvature $\Omega$, initial center $i_0$ and displacement of the trap center $\Delta x$. In all our calculations, the shift $\Delta x$ is set to be 5, close to that in the experiment~\cite{Strohmaier2007}. For brevity we work in units where the lattice spacing is unity hereafter. To study the dipolar motion, we are interested in the position of the COM, which is defined as
\begin{equation}
  X=\frac{\sum_{i,\sigma}in_{i,\sigma}}{2N}-i_0.
\end{equation}
Here $N_\uparrow\!=\!N_\downarrow\!=\!N$ is the particle number per ``spin'' species of a population balanced gas.

\section{Attractive Interactions}
\subsection{Weakly interacting regime: underdamped oscillation} 
Unlike the bosonic case, the fermionic dipole oscillations are weakly damped even in the non-interacting case. This damping is caused by the dephasing of particles in the same ``spin'' state due to Pauli's principle. When a weak on-site interaction is introduced, the dephasing will be more pronounced because interaction between particles with different spins also contributes to incoherence.  

\begin{figure}[ht]
  \begin{center}    
    \begin{tabular}{c}    
      \resizebox{0.45\textwidth}{!}{\includegraphics{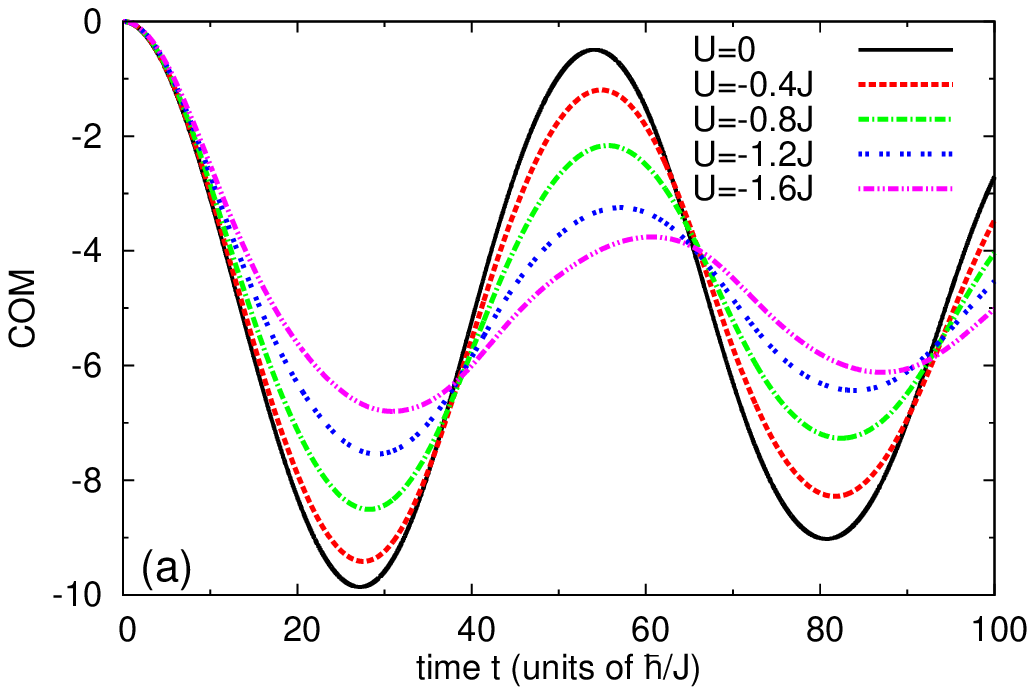}} \\
      \resizebox{0.45\textwidth}{!}{\includegraphics{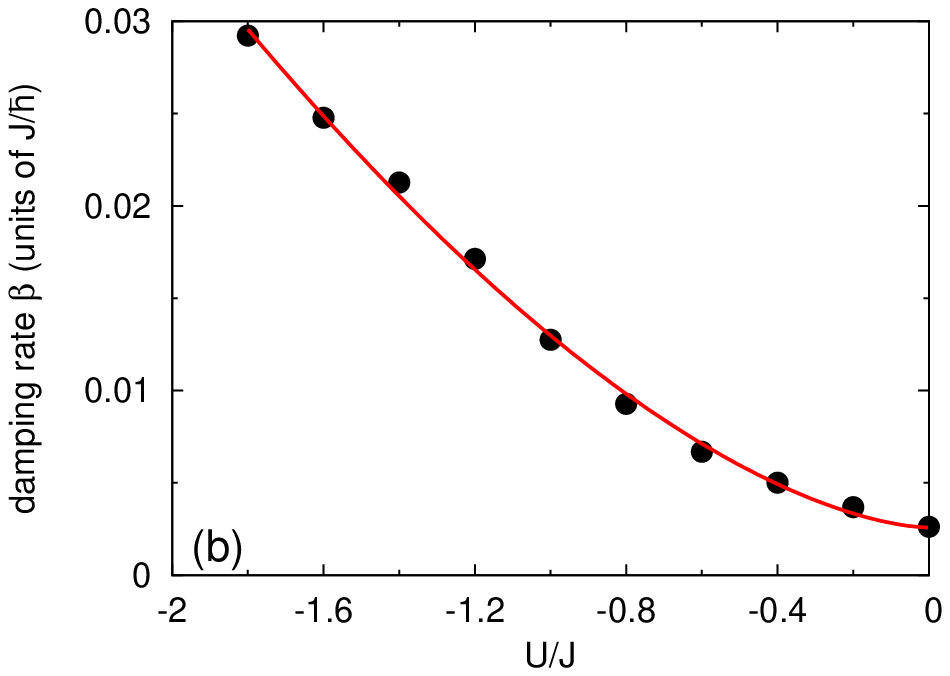}} 
    \end{tabular}
\end{center} 
\caption{\label{damping}(Color online) (a) The position of COM as a function of time with different interactions. (b) $U/J$ dependence of the damping rate $\beta$. The line is a fit of the data to Eq.~\ref{eqn:fit}. The trap curvature is $\Omega\!=\!0.005J$, equivalent to a trapping frequency of about $50$ Hz in the experiment.}
\end{figure}

Fig.~\ref{damping} (a) shows these oscillations in a system with $L\!=\!80$ and $N\!=\!16$. In this regime the COM motion can be described by an underdamped harmonic oscillation around the new trap center after the shift. We fit the motion of COM to
\begin{equation}
  X(t)=A+Be^{-\beta t}\cos\omega t,  
\end{equation}
where $\beta$ is the damping rate and $\omega$ is the frequency. Fig.~\ref{damping} (b) shows the damping rate $\beta$ as a function of the interaction strength $U/J$. The results are well fitted by the empirical power law
\begin{equation}\label{eqn:fit}
  \frac{\beta}{J/\hbar}=a\left(\frac{|U|}{J}\right)^{b}+c,
\end{equation}
with $a\!=\!0.0104$, $b\!=\!1.62$, and $c\!=\!0.00258$. The first term is a power law describing the interaction effect and the second term describes the dephasing due to Pauli's principle. 

\subsection{Strongly interacting regime: overdamped relaxation}

In the strongly attractive case the experiment~\cite{Strohmaier2007} shows a relaxational COM motion. Here we study this overdamped behavior in a system with $L\!=\!80$ and $\Omega\!=\!0.005J$. First, we find a critical value $U_{\text{c}}\!\approx\!-3.5J$ for the {\it crossover} between underdamped oscillations and overdamped relaxation in simulations with $N$=10--20 particles per ``spin'' species. Note that this critical value is estimated by examining whether the COM oscillates around the new trap center or not after the shift.
\begin{figure}[ht]
  \begin{center}    
    \begin{tabular}{c}    
      \resizebox{0.45\textwidth}{!}{\includegraphics{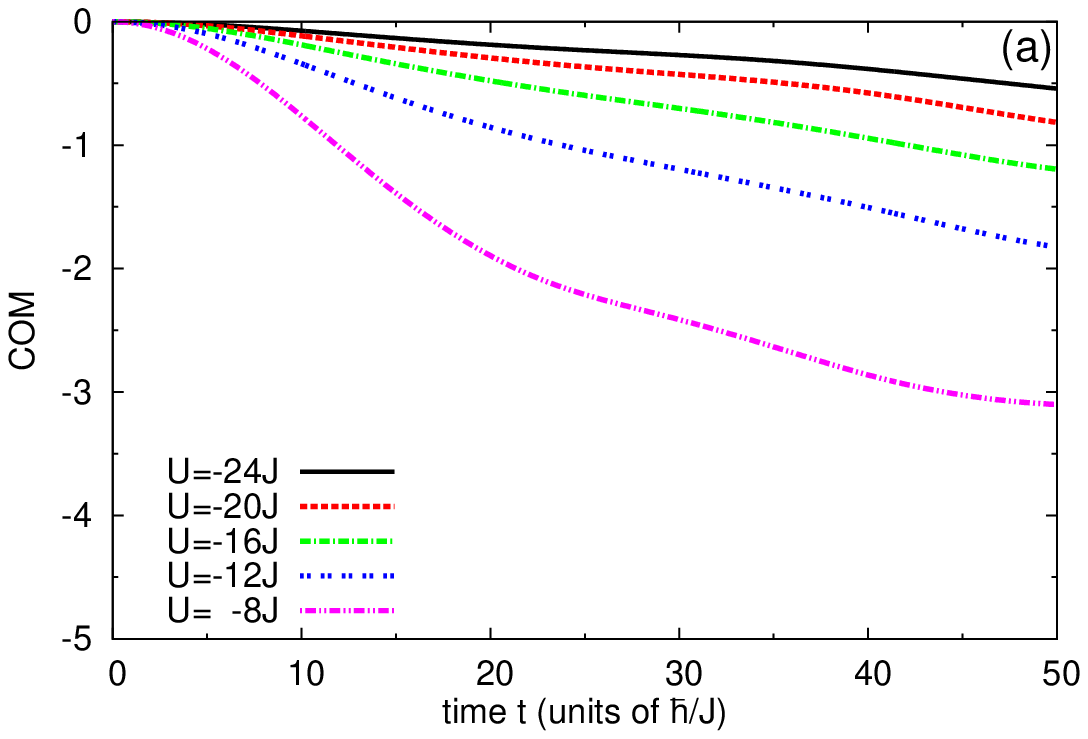}} \\
      \resizebox{0.45\textwidth}{!}{\includegraphics{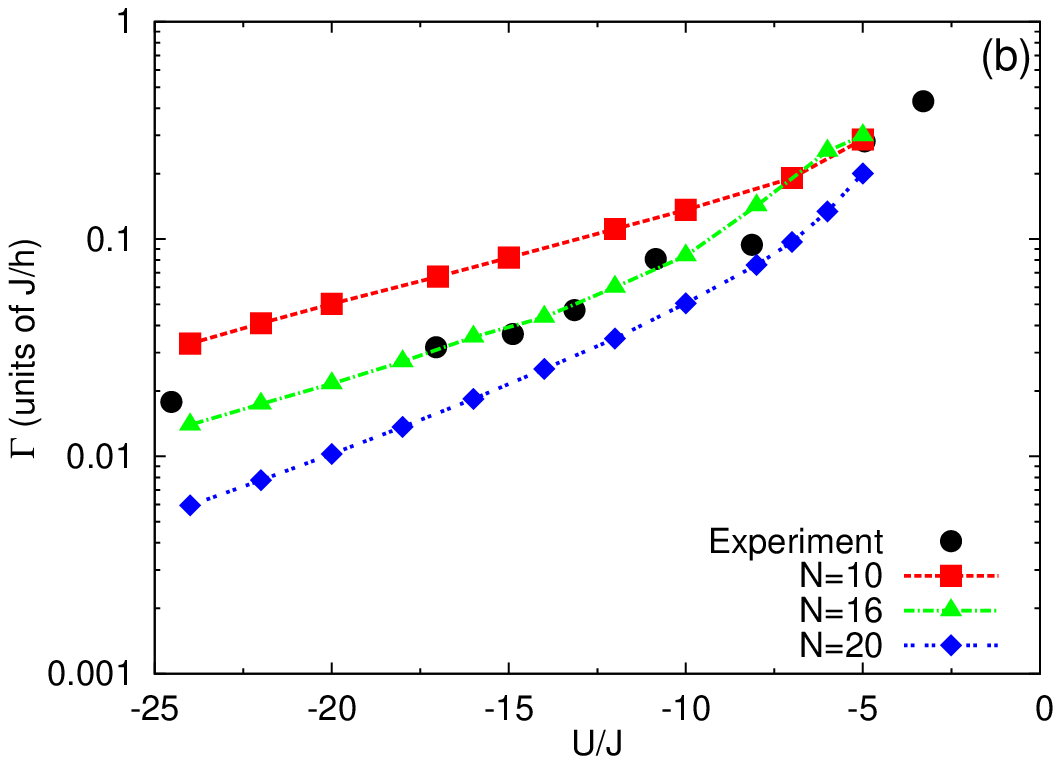}} 
    \end{tabular}
  \end{center} 
  \caption{\label{relaxation}(Color online) (a) COM position as a function of time in a system with $N=16$. The interaction strengths $U$ are indicated in the figure. (b) The relaxation rate $\Gamma$ vs the ratio $U/J$ for various particle numbers per ``spin'' in the lattice. The experimental (black circle) results are obtained from Fig.~3 of Ref.~\cite{Strohmaier2007}. Note that our results can be qualitatively compared with the experimental data although the experiment was performed in a three-dimensional optical lattice.}
\end{figure}

As can be seen in Fig.~\ref{relaxation} (a), the motion of the COM in the strong interacting regime is characterized by a relaxation towards the new trap center. Unlike the previous weakly interacting regime, the COM mass can no longer oscillate around the new trap center, although its motion still shows some negligible undulations (see the case $U\!=\!-8J$ in Fig.~\ref{relaxation} (a) for example). As the attraction increases the fermions form stronger bound pairs with a larger effective mass. This slows the motion of the atoms towards the new trap center. 

We extract the relaxation rate by fitting the curves to an exponential decay
\begin{equation}
  z(t)=z_{\infty}(1-e^{-\Gamma t}),
\end{equation}
where $z_{\infty}$ is the position at $t\rightarrow\infty$, which is assumed to be the new trap center (in our case, $z_{\infty}\!=\!-5$). 

The relaxation rate as a function of interaction for different particle numbers per ``spin'' is shown in Fig.~\ref{relaxation} (b). We find very good agreement with experimental data in the regime of strong interactions. $N\!\!=\!\!16$ corresponds to a system with half filling at the center, consistent with the experiment. We also find that for a given interaction strength $U$, the relaxation rate decreases as the particle number increases, as has been observed experimentally. The good agreement is a strong evidence that the single-band Hubbard model can capture all the features of the experiment. 

In the strong interaction limit the fermions form local singlet pairs on a lattice site. The transport properties in this regime are thus governed by the dynamics of these local pairs, which can be approximated as hard-core bosons (HCB). Here we perform simulations to test the validity of this mapping to an effective HCB model 
\begin{multline}
  \hat{H}_B(t)=-J_{\text{eff}}\sum_{i}(\hat{b}_{i+1}^\dag\hat{b}^{}_{i}+ \text{H.c.})\\
  +\Omega'\sum_{i}\hat{b}_{i}^\dag\hat{b}^{}_i(i-i_0-\Theta(t)\Delta x)^2,
\end{multline}
where $\hat{b}_i$'s are bosonic annihilation operators with additional hard-core constrains $\hat{b}^{\dag 2}_i\!=\!\hat{b}^2_i\!=\!0$ and $\{\hat{b}^{}_i,\hat{b}^\dag_i\}\!=\!1$. The effective hopping integral of a pair is 
\begin{equation}
  J_{\text{eff}}=\frac{\sqrt{16J^2+U^2}}{4}-\frac{1}{4}|U|,
\end{equation}
which reduces to $2J^2/|U|$ in the large $|U|$ limit~\cite{Micnas1990}. Since one HCB represents a pair of fermions, whose bare mass is twice that of a fermion, the trapping curvature $\Omega'\!=\!2\Omega$ is twice that of the original one. 

By exactly diagonalizing the model it is straightforward to obtain numerically exact results for the time evolution of the system. 

\begin{figure}[ht]
  \begin{center}
    \begin{tabular}{cc}    
      \resizebox{0.25\textwidth}{!}{\includegraphics{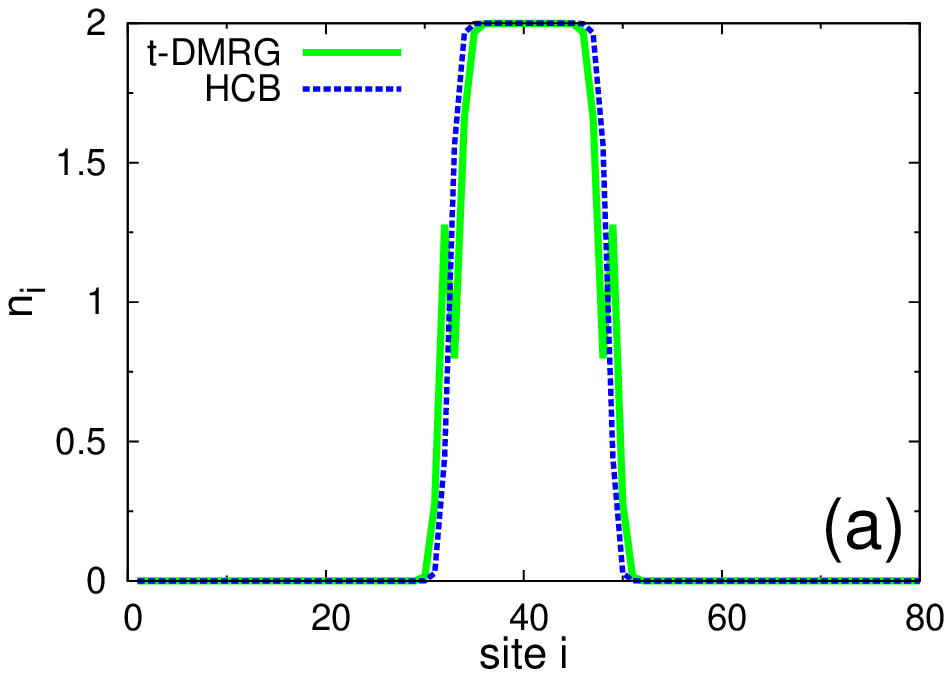}} & 
      \resizebox{0.25\textwidth}{!}{\includegraphics{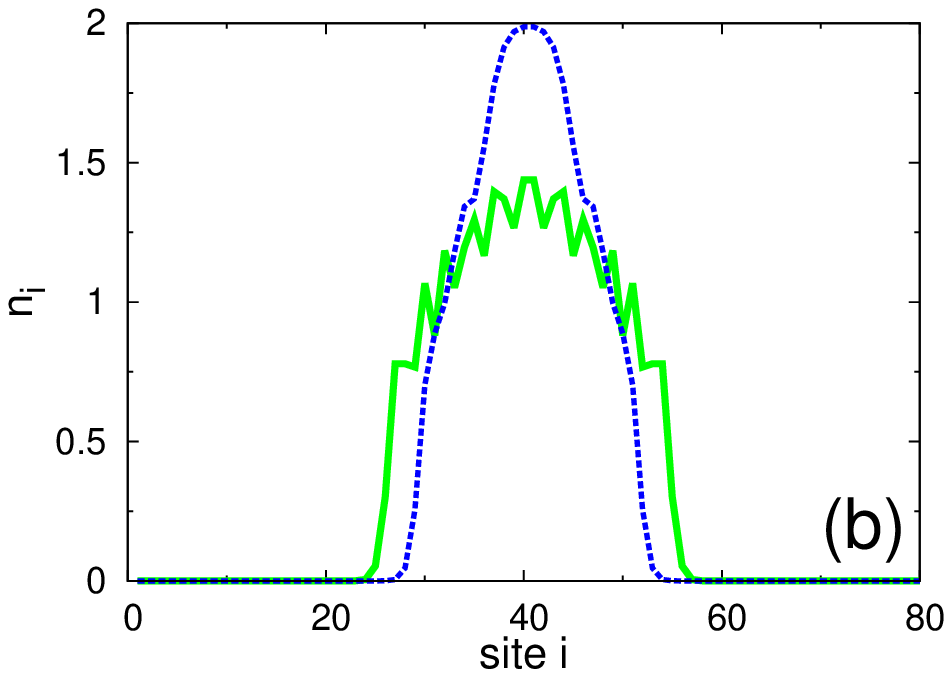}}  
    \end{tabular}
    \begin{tabular}{c}    
      \resizebox{0.45\textwidth}{!}{\includegraphics{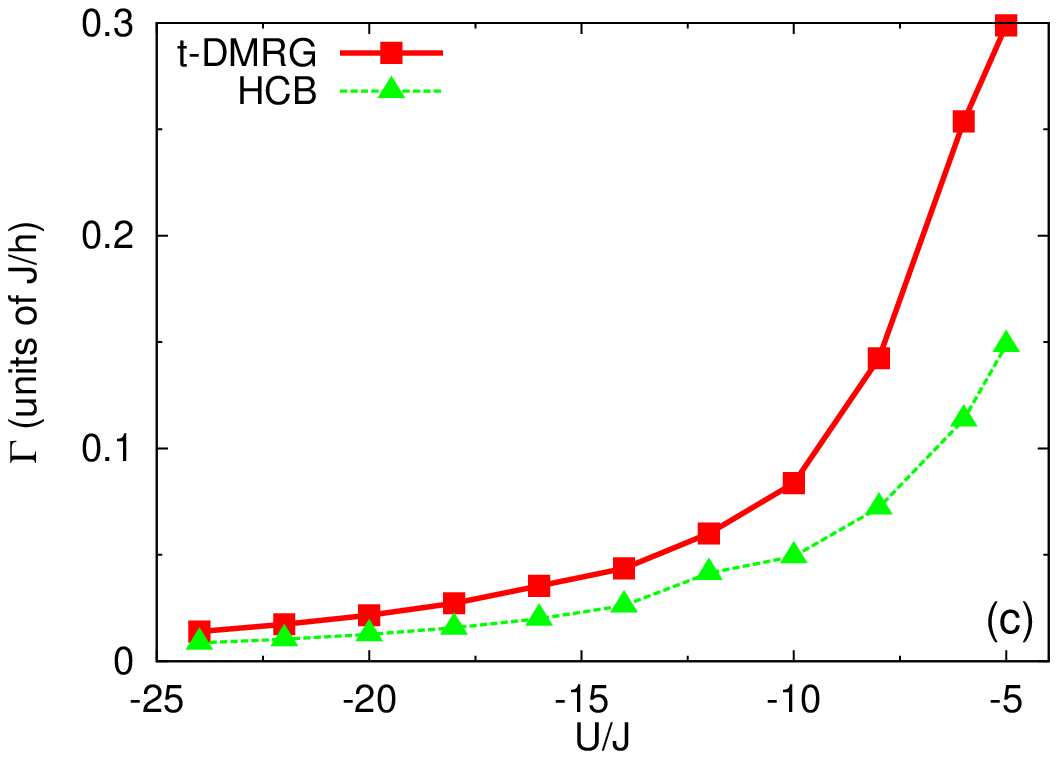}} 
    \end{tabular}
  \end{center} 
  \caption{\label{mapping}(Color online) Static and dynamical mappings to the HCB model. Density distributions in the initial states for (a) $U=-24J$ and (b) $U=-5J$, respectively. (c) The relaxation rate $\Gamma$ as a function of $U/J$.  The results of both the original fermionic model and the HCB model are shown for comparison. Here we use $N=16$ particles per ``spin'' species.}
\end{figure}

To test the validity of the HCB approximation for static properties, we  consider  the density distribution. From the comparison of the fermionic and the effective model shown in Fig.~\ref{mapping} (a) and (b), we find that the mapping of static quantities to an HCB model is valid in the large $|U|$ limit, but less so in the regime of intermediate interactions. The HCB model overestimates the central density at $U\!\!\sim\!\!-5J$ because it assumes most fermions form fully local pairs even when the attractive interaction is not strong enough. For the dynamical mapping, the HCB model captures the interaction dependence of $\Gamma$ qualitatively. However, the poor mapping of static quantities at $U\!\!\sim\!\!-5J$ significantly compromises the approximate relaxation rate $\Gamma$ based on the HCB model, which deviates from the exact results by about 50\%. The HCB model also underestimates the relaxation rate in the intermediate regime, where single particles and local pairs coexist. 

We can improve the approximation by using an extended HCB Hamiltonian
\begin{equation}
  \hat{H}'_B(t)=\hat{H}_B(t)+V\sum_{i}\hat{n}_i\hat{n}_{i+1},
\end{equation}
where the term with $V>0$ describes the nearest neighbor interaction between local pairs. This repulsive term will reduce the deviation in density profiles shown in Fig.~\ref{mapping} (b). However, the dynamical mapping cannot be perfect even with this extended HCB model, because the internal dynamics of the pairs are not taken into account. 

\section{Repulsive Interactions}
In the repulsive case we need to distinguish between low and high density regimes. At low density (or in a shallow trap at fixed particle number), the fermions are delocalized, leading to compressible metallic states with central density less than unity at $U=0$. In this case, Mott-insulating states will never appear even when $U$ is increased due to the low density. On the other hand, if the particle density is high enough (or the external trapping potential deep enough), a central Mott domain will be formed beyond some critical value of $U$. 

\begin{figure}[ht]
  \begin{center}    
    \begin{tabular}{c}
      \resizebox{0.45\textwidth}{!}{\includegraphics{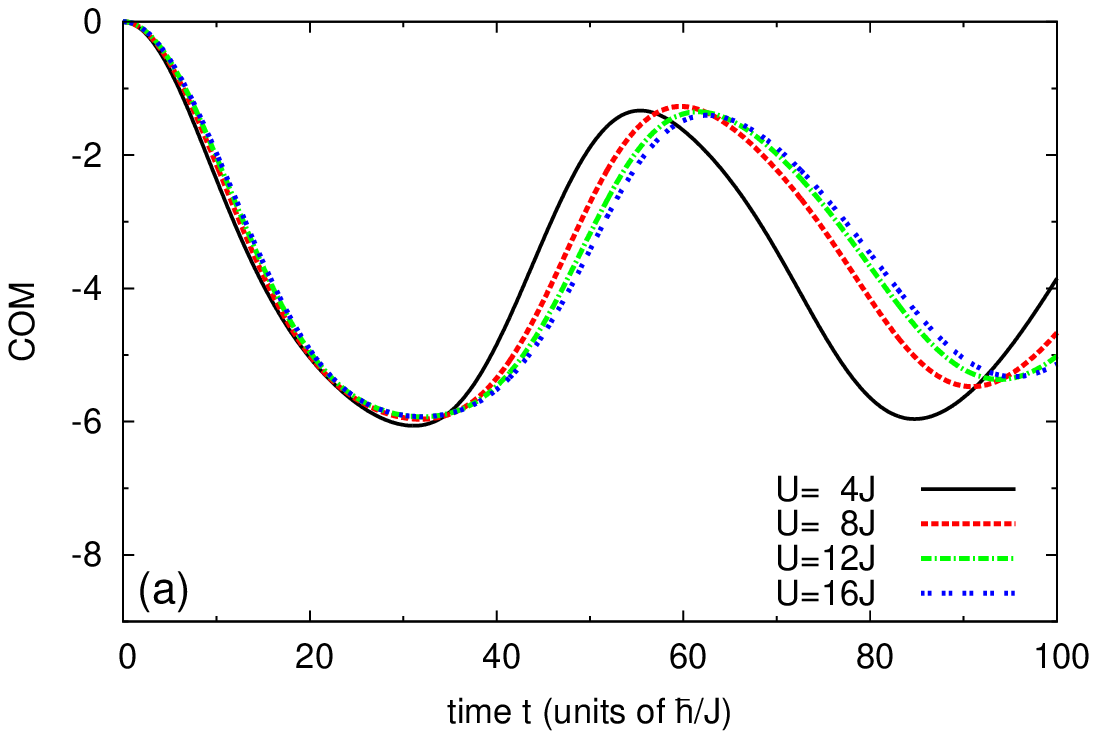}} \\
      \resizebox{0.45\textwidth}{!}{\includegraphics{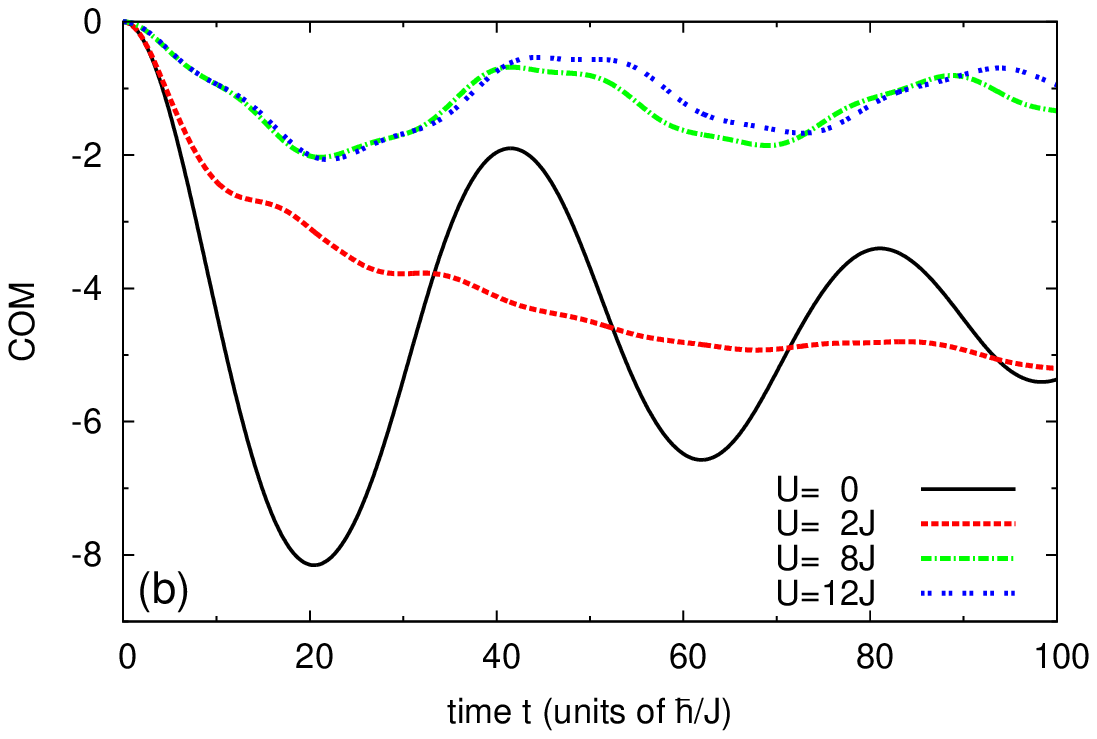}} 
    \end{tabular}
  \end{center} 
  \caption{\label{transition}(Color online) COM motions for (a) $\Omega=0.005J$ and (b) $\Omega=0.01J$, respectively. In each panel, the on-site interactions are indicated in the figure. Other parameters are: $L=80$ and $N=16$.}
\end{figure}

In the low density (shallow trap) case, the COM motion is always underdamped, as shown in Fig.~\ref{transition} (a) and converges in the large $U$ limit.

\begin{figure}[ht]
  \begin{center}    
    \begin{tabular}{c}
      \resizebox{0.45\textwidth}{!}{\includegraphics{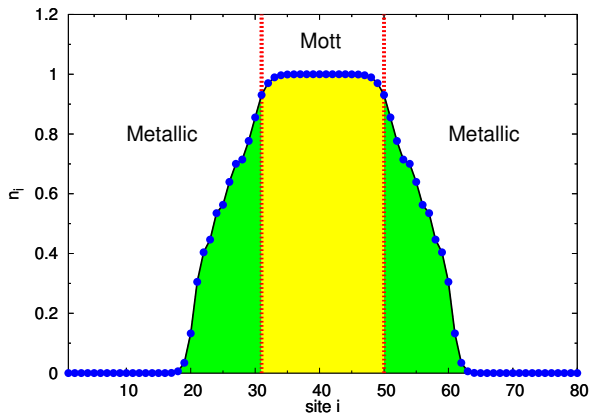}} 
    \end{tabular}
    \end{center} 
  \caption{\label{density}(Color online) Density distribution for $U=12J$. The particle number in the Mott domain is about 16, and there are about 8 fermions in each metallic domain on the left and right. Therefore, due to the imcompressibility of the central Mott phase, only a quarter of all the particles contribute to the oscillation after the sudden shift.}
\end{figure}

However, for a high density (deep trap), the situation is very different, as shown in Fig.~\ref{transition} (b). Here, there is a crossover between underdamped and overdamped behaviour at around $U\!=\!2J$. However, when $U$ is further increased beyond $4J$, a Mott domain will be formed at the center (see Fig.~\ref{density} for example), and oscillations reappear on the background of a very slow relaxation. By increasing $U$ the dipolar motion exhibits a collapse and revival of oscillation with interaction strength.

Analyzing the revival of the COM oscillations for $\Omega\!=\!0.01J$ and $U\!=\!12J$ as shown in Fig.~\ref{density}, we see that their frequency is the same as the non-interacting one, {\it i.e.}, the harmonic trap, similar to the case of cyclotron motion of electrons \cite{Kohn1961}. Furthermore, the oscillating amplitude is about one quarter of the non-interacting case, suggesting that only one quarter of the atoms are oscillating. This agrees with the corresponding density profile (see Fig.~\ref{density}), which shows that half of the atoms form the Mott plateau, leading to one quarter at the each edge of the domain. We argue that the whole COM motion can be interpreted within a two-fluid model as follows: The metallic component at the edges of the Mott domain oscillates with moderate damping, while the incompressible Mott-insulating counterpart becomes nearly localized. In this two-fluid model, the oscillation due to the metallic component is also conceptually equivalent to the residual current in the effective model with two coupled bands separated by $U$ \cite{Montangero2009}. 

\section{Conclusion}
Using tDMRG, we have presented an accurate analysis of the dipolar motion of an interacting Fermi gas in an optical lattice. In the regime of attractive interactions, the numerical calculations for the COM dynamics after a sudden displacement of the trap minimum are in good agreement with experimental results. 

To further theoretical understanding, we have mapped the fermionic model to an effective HCB model in the strong interaction limit. To improve the validity of the mapping for intermediate interactions, an extended HCB model is necessary and will be subject of further studies. 

For repulsive interactions a very different behaviour has been found. In a deep trap, we find a revival of the oscillation for the COM by increasing the interaction strength when the trap is deep enough to support a central Mott domain with increasing $U$. This revival is due to the fact that as the central Mott plateau is nearly frozen, there will inevitably mobile metallic states at the edges. By comparing with the lower density case, we conclude that the overdamped relaxation in the repulsive Hubbard model can be regarded as a dynamical signature for the emergence of the Mott domain for an inhomogeneous Fermi gas in an optical lattice.

\begin{acknowledgments}
We acknowledge part of financial support from HKSAR RGC Grant No.~701009. U.S. acknowledges support by DFG. M.T. acknowledges support of the Aspen Center for Physics under NSF grant 1066293.
\end{acknowledgments}

\bibliography{library}
\bibliographystyle{apsrev4-1}

\end{document}